
\input phyzzx
\catcode`@=11 
\def\space@ver#1{\let\@sf=\empty \ifmmode #1\else \ifhmode
   \edef\@sf{\spacefactor=\the\spacefactor}\unskip${}#1$\relax\fi\fi}
\def\attach#1{\space@ver{\strut^{\mkern 2mu #1} }\@sf\ }
\newtoks\foottokens
\newbox\leftpage \newdimen\fullhsize \newdimen\hstitle \newdimen\hsbody
\newif\ifreduce  \reducefalse
\def\almostshipout#1{\if L\lr \count2=1
      \global\setbox\leftpage=#1 \global\let\lr=R
  \else \count2=2
    \shipout\vbox{\special{dvitops: landscape}
      \hbox to\fullhsize{\box\leftpage\hfil#1}} \global\let\lr=L\fi}
\def\smallsize{\relax
\font\eightrm=cmr8 \font\eightbf=cmbx8 \font\eighti=cmmi8
\font\eightsy=cmsy8 \font\eightsl=cmsl8 \font\eightit=cmti8
\font\eightt=cmtt8
\def\eightpoint{\relax
\textfont0=\eightrm  \scriptfont0=\sixrm
\scriptscriptfont0=\sixrm
\def\rm{\fam0 \eightrm \f@ntkey=0}\relax
\textfont1=\eighti  \scriptfont1=\sixi
\scriptscriptfont1=\sixi
\def\oldstyle{\fam1 \eighti \f@ntkey=1}\relax
\textfont2=\eightsy  \scriptfont2=\sixsy
\scriptscriptfont2=\sixsy
\textfont3=\tenex  \scriptfont3=\tenex
\scriptscriptfont3=\tenex
\def\it{\fam\itfam \eightit \f@ntkey=4 }\textfont\itfam=\eightit
\def\sl{\fam\slfam \eightsl \f@ntkey=5 }\textfont\slfam=\eightsl
\def\bf{\fam\bffam \eightbf \f@ntkey=6 }\textfont\bffam=\eightbf
\scriptfont\bffam=\sixbf   \scriptscriptfont\bffam=\sixbf
\def\tt{\fam\ttfam \eightt \f@ntkey=7 }
\def\caps{\fam\cpfam \tencp \f@ntkey=8 }\textfont\cpfam=\tencp
\setbox\strutbox=\hbox{\vrule height 7.35pt depth 3.02pt width\z@}
\samef@nt}
\def\Eightpoint{\eightpoint \relax
  \ifsingl@\subspaces@t2:5;\else\subspaces@t3:5;\fi
  \ifdoubl@ \multiply\baselineskip by 5
            \divide\baselineskip by 4\fi }
\parindent=16.67pt
\itemsize=25pt
\thinmuskip=2.5mu
\medmuskip=3.33mu plus 1.67mu minus 3.33mu
\thickmuskip=4.17mu plus 4.17mu
\def\thinspace{\kern .13889em }
\def\negthinspace{\kern-.13889em }
\def\enspace{\kern.416667em }
\def\enskip{\hskip.416667em\relax}
\def\quad{\hskip.83333em\relax}
\def\qquad{\hskip1.66667em\relax}
\def\crr{\cropen{8.3333pt}}
\foottokens={\Eightpoint\singlespace}
\def\papersize{\SIZE\OFFSET\skip\footins=\bigskipamount}
\def\SIZE{\hsize=11.8truecm\vsize=17.5truecm}
\def\OFFSET{\voffset=-1.3truecm\hoffset=  .14truecm}
\message{STANDARD CERN-PREPRINT FORMAT}
\def\attach##1{\space@ver{\strut^{\mkern 1.6667mu ##1} }\@sf\ }
\def\PH@SR@V{\doubl@true\baselineskip=20.08pt plus .1667pt minus .0833pt
             \parskip = 2.5pt plus 1.6667pt minus .8333pt }
\def\author##1{\vskip\frontpageskip\titlestyle{\tencp ##1}\nobreak}
\def\address##1{\par\kern 4.16667pt\titlestyle{\tenpoint\it ##1}}
\def\andaddress{\par\kern 4.16667pt \centerline{\sl and} \address}
\def\abstract{\vskip2\frontpageskip\centerline{\tenrm Abstract}
              \vskip\headskip }
\def\cases##1{\left\{\,\vcenter{\Tenpoint\m@th
    \ialign{$####\hfil$&\quad####\hfil\crcr##1\crcr}}\right.}
\def\matrix##1{\,\vcenter{\Tenpoint\m@th
    \ialign{\hfil$####$\hfil&&\quad\hfil$####$\hfil\crcr
      \mathstrut\crcr\noalign{\kern-\baselineskip}
     ##1\crcr\mathstrut\crcr\noalign{\kern-\baselineskip}}}\,}
\Tenpoint
}
\def\Smallsize{\smallsize\reducetrue
\let\lr=L
\hstitle=8truein\hsbody=4.75truein\fullhsize=24.6truecm\hsize=\hsbody
\output={
  \almostshipout{\leftline{\vbox{\makeheadline
  \pagebody\makefootline}}}\advancepageno
     }
\special{dvitops: landscape}
\def\makeheadline{
\iffrontpage\line{\the\headline}
             \else\vskip .0truecm\line{\the\headline}\vskip .5truecm \fi}
\def\makefootline{\iffrontpage\vskip  0.truecm\line{\the\footline}
               \vskip -.15truecm\line{\the\date\hfil}
              \else\line{\the\footline}\fi}
\paperheadline={
\iffrontpage\hfil
               \else
               \tenrm\hss $-$\ \folio\ $-$\hss\fi    }
\paperstyle}
%
%
%
%
%
%
%
%
%
\newcount\referencecount     \referencecount=0
\newif\ifreferenceopen       \newwrite\referencewrite
\newtoks\rw@toks
\def\NPrefmark#1{\attach{\scriptscriptstyle [ #1 ] }}
\let\PRrefmark=\attach
\def\refmark#1{\relax\ifPhysRev\PRrefmark{#1}\else\NPrefmark{#1}\fi}
\def\refend{\refmark{\number\referencecount}}
\newcount\lastrefsbegincount \lastrefsbegincount=0
\def\refsend{\refmark{\count255=\referencecount
   \advance\count255 by-\lastrefsbegincount
   \ifcase\count255 \number\referencecount
   \or \number\lastrefsbegincount,\number\referencecount
   \else \number\lastrefsbegincount-\number\referencecount \fi}}
\def\refch@ck{\chardef\rw@write=\referencewrite
   \ifreferenceopen \else \referenceopentrue
   \immediate\openout\referencewrite=referenc.texauxil \fi}
%
{\catcode`\^^M=\active 
  \gdef\obeyendofline{\catcode`\^^M\active \let^^M\ }}%
%
{\catcode`\^^M=\active 
  \gdef\ignoreendofline{\catcode`\^^M=5}}
{\obeyendofline\gdef\rw@start#1{\def\t@st{#1} \ifx\t@st\blankend%
\endgroup \@sf \relax \else \ifx\t@st\bl@nkend \endgroup \@sf \relax%
\else \rw@begin#1
\backtotext
\fi \fi } }
{\obeyendofline\gdef\rw@begin#1
{\def\n@xt{#1}\rw@toks={#1}\relax%
\rw@next}}
\def\blankend{}
{\obeylines\gdef\bl@nkend{
}}
\newif\iffirstrefline  \firstreflinetrue
\def\rwr@teswitch{\ifx\n@xt\blankend \let\n@xt=\rw@begin %
 \else\iffirstrefline \global\firstreflinefalse%
\immediate\write\rw@write{\noexpand\obeyendofline \the\rw@toks}%
\let\n@xt=\rw@begin%
      \else\ifx\n@xt\rw@@d \def\n@xt{\immediate\write\rw@write{%
        \noexpand\ignoreendofline}\endgroup \@sf}%
             \else \immediate\write\rw@write{\the\rw@toks}%
             \let\n@xt=\rw@begin\fi\fi \fi}
\def\rw@next{\rwr@teswitch\n@xt}
\def\rw@@d{\backtotext} \let\rw@end=\relax
\let\backtotext=\relax

\newdimen\refindent     \refindent=30pt
\def\refitem#1{\par \hangafter=0 \hangindent=\refindent \Textindent{#1}}
\def\REFNUM#1{\space@ver{}\refch@ck \firstreflinetrue%
 \global\advance\referencecount by 1 \xdef#1{\the\referencecount}}
\def\refnum#1{\space@ver{}\refch@ck \firstreflinetrue%
 \global\advance\referencecount by 1 \xdef#1{\the\referencecount}\refend}

\def\REF#1{\REFNUM#1%
 \immediate\write\referencewrite{%
 \noexpand\refitem{#1.}}%
\begingroup\obeyendofline\rw@start}
\def\ref{\refnum\?%
 \immediate\write\referencewrite{\noexpand\refitem{\?.}}%
\begingroup\obeyendofline\rw@start}
\def\Ref#1{\refnum#1%
 \immediate\write\referencewrite{\noexpand\refitem{#1.}}%
\begingroup\obeyendofline\rw@start}
\def\REFS#1{\REFNUM#1\global\lastrefsbegincount=\referencecount
\immediate\write\referencewrite{\noexpand\refitem{#1.}}%
\begingroup\obeyendofline\rw@start}
\newif\ifmref  
\newif\iffref  
\def\xrefsend{\xrefmark{\count255=\referencecount
\advance\count255 by-\lastrefsbegincount
\ifcase\count255 \number\referencecount
\or \number\lastrefsbegincount,\number\referencecount
\else \number\lastrefsbegincount-\number\referencecount \fi}}
\def\xrefsdub{\xrefmark{\count255=\referencecount
\advance\count255 by-\lastrefsbegincount
\ifcase\count255 \number\referencecount
\or \number\lastrefsbegincount,\number\referencecount
\else \number\lastrefsbegincount,\number\referencecount \fi}}
\def\xREFNUM#1{\space@ver{}\refch@ck\firstreflinetrue%
\global\advance\referencecount by 1
\xdef#1{\xrefend}}
\def\xrefend{\xrefmark{\number\referencecount}}
\def\xrefmark#1{[{#1}]}
\def\xRef#1{\xREFNUM#1\immediate\write\referencewrite%
{\noexpand\refitem{#1 }}\begingroup\obeyendofline\rw@start}%
\def\xREFS#1{\xREFNUM#1\global\lastrefsbegincount=\referencecount%
\immediate\write\referencewrite{\noexpand\refitem{#1 }}%
\begingroup\obeyendofline\rw@start}
\def\rrr#1#2{\relax\ifmref{\iffref\xREFS#1{#2}%
\else\xRef#1{#2}\fi}\else\xRef#1{#2}\xrefend\fi}
\def\multref#1#2{\mreftrue\freftrue{#1}%
\freffalse{#2}\mreffalse\xrefsend}
\referencecount=0
\def\refbreak{\hfil\penalty200\hfilneg}
\def\paperstyle{\papers}
\paperstyle   
%
%
%
\def\slacpub{\afterassignment\slacp@b\toks@}
\def\slacp@b{\edef\n@xt{\Pubnum={NIKHEF--H/\the\toks@}}\n@xt}
\let\pubnum=\slacpub
\expandafter\ifx\csname eightrm\endcsname\relax
    \let\eightrm=\ninerm \let\eightbf=\ninebf \fi

\font\seventeencp=cmcsc10 scaled\magstep3

\newif\ifCONF \CONFfalse
\newif\ifBREAK \BREAKfalse
\newif\ifsectionskip \sectionskiptrue

%
%
%
%
\def\NuclPhysProc{
\let\lr=L
\hstitle=8truein\hsbody=4.75truein\fullhsize=21.5truecm\hsize=\hsbody
\hstitle=8truein\hsbody=4.75truein\fullhsize=20.7truecm\hsize=\hsbody
\output={
  \almostshipout{\leftline{\vbox{\makeheadline
  \pagebody\makefootline}}}\advancepageno
     }
\def\papersize{\SIZE\OFFSET\skip\footins=\bigskipamount}
\def\SIZE{\hsize=10.0truecm\vsize=27.0truecm}
\def\OFFSET{\voffset=-1.4truecm\hoffset=-2.40truecm}
\message{NUCLEAR PHYSICS PROCEEDINGS FORMAT}
\def\makeheadline{
\iffrontpage\line{\the\headline}
             \else\vskip .0truecm\line{\the\headline}\vskip .5truecm \fi}
\def\makefootline{\iffrontpage\vskip  0.truecm\line{\the\footline}
               \vskip -.15truecm\line{\the\date\hfil}
              \else\line{\the\footline}\fi}
\paperheadline={\hfil}
\paperstyle}
%
%
%
%

%
%
%
%

%
%
%
%
\def\ReprintVolume{\smallsize
\def\papersize{\hsize=18.0truecm\vsize=25.1truecm\voffset -.73truecm
    \hoffset -.65truecm\skip\footins=\bigskipamount
    \normaldisplayskip= 20pt plus 5pt minus 10pt}
\message{REPRINT VOLUME FORMAT}
\paperstyle\baselineskip=.425truecm\parskip=0truecm
\def\makeheadline{
\iffrontpage\line{\the\headline}
             \else\vskip .0truecm\line{\the\headline}\vskip .5truecm \fi}
\def\makefootline{\iffrontpage\vskip  0.truecm\line{\the\footline}
               \vskip -.15truecm\line{\the\date\hfil}
              \else\line{\the\footline}\fi}
\paperheadline={
\iffrontpage\hfil
               \else
               \tenrm\hss $-$\ \folio\ $-$\hss\fi    }
\def\sectionfont{\bf}    }
%
%
%
%
\def\SIZE{\hsize=15.73truecm\vsize=23.11truecm}
\def\OFFSET{\voffset=0.0truecm\hoffset=0.truecm}
\message{DEFAULT FORMAT}
\def\papersize{\SIZE\OFFSET\skip\footins=\bigskipamount
\normaldisplayskip= 35pt plus 3pt minus 7pt}
\def\CERN{\address{{\sl CERN, 1211 Geneva 23, Switzerland\
\phantom{XX}\ }}}
\Pubnum={\rm NIKHEF--H/\the\pubnum }
\def\title#1{\vskip\frontpageskip\vskip .50truein
     \titlestyle{\seventeencp #1} \vskip\headskip\vskip\frontpageskip
     \vskip .2truein}
\def\author#1{\vskip .27truein\titlestyle{#1}\nobreak}
\def\andauthor{\vskip .27truein\centerline{and}\author}
\def\p@bblock{\begingroup \tabskip=\hsize minus \hsize
   \baselineskip=1.5\ht\strutbox \topspace+2\baselineskip
   \halign to\hsize{\strut ##\hfil\tabskip=0pt\crcr
  \the \Pubnum\cr}\endgroup}
\def\makefootline{\iffrontpage\vskip .27truein\line{\the\footline}
                 \vskip -.1truein
              \else\line{\the\footline}\fi}
\paperfootline={\iffrontpage
\hfil\else\hfil\fi}

\def\abstract{\vskip2\frontpageskip\centerline{\twelvebf Abstract}
              \vskip\headskip }

\paperheadline={
\iffrontpage\hfil
               \else
               \twelverm\hss $-$\ \folio\ $-$\hss\fi}
%
%
\def\nup#1({\refbreak\ Nucl.\ Phys.\ $\underline {B#1}$\ (}
\def\plt#1({\refbreak\ Phys.\ Lett.\ $\underline  {#1}$\ (}
\def\cmp#1({\refbreak\ Commun.\ Math.\ Phys.\ $\underline  {#1}$\ (}
\def\prp#1({\refbreak\ Physics\ Reports\ $\underline  {#1}$\ (}
\def\prl#1({\refbreak\ Phys.\ Rev.\ Lett.\ $\underline  {#1}$\ (}
\def\prv#1({\refbreak\ Phys.\ Rev. $\underline  {D#1}$\ (}
\def\und#1({            $\underline  {#1}$\ (}
%
%

\def\rB{\hfil\penalty1000\hfilneg}
%
%
\hyphenation{sym-met-ric anti-sym-me-tric re-pa-ra-me-tri-za-tion
Lo-rentz-ian a-no-ma-ly di-men-sio-nal two-di-men-sio-nal}
%
%
%
%

\def\coeff#1#2{{\textstyle { #1 \over #2}}\displaystyle}
\def\boxit#1{\vbox{\hrule\hbox{\vrule\kern3pt
\vbox{\kern3pt#1\kern3pt}\kern3pt\vrule}\hrule}}
\message{ by V.K, W.L and A.S}
\catcode`@=12
\paperstyle
\def\Vaf {\rrr\Vaf {C.~Vafa, \nup273 (1986) 592.}}
\def\VafN{\rrr\VafN{C.~Vafa,
Mod.~Phys.~Lett.~\und{A4} (1989) 1169.}}
\def\ScYd{\rrr\ScYd{A.N~Schellekens and S.~Yankielowicz,
\nup330 (1990) 103;\ \rB Tables Supplement
CERN-TH.5440T/89 (1989)   (unpublished).}}
\def\ScYg{\rrr\ScYg{A.N.~Schellekens and S.~Yankielowicz,
Int.~J.~Mod.~ Phys.~\und{A5} (1990) 2903.}}
\def\Vafa{\rrr\Vafa{C.~Vafa,
\plt 206B (1988) 421.}}
\def\Scha{\rrr\Scha{
A.N.~Schellekens,
\plt B244 (1990) 255.}}
\def\ScYb{\rrr\ScYb{
A.N.~Schellekens and S.~Yankielowicz,
\plt B227 (1989) 387.}}
\def\ScYa{\rrr\ScYa{
A.N.~Schellekens and S.~Yankielowicz,
\nup 327 (1989) 673.}}
\def\Alig{\rrr\Alig{K.~Intriligator,
\nup 332 (1990) 541.}}
\def\FKSS{\rrr\FKSS{J.~Fuchs, A.~Klemm, C.~Scheich and M.~Schmidt,
\plt B232 (1989) 317; Ann.~Phys.\und{204} (1990) 1.}}
\def\BeBe{\rrr\BeBe {B.~Gato-Rivera and A.N.~Schellekens,
\nup 353 (1991) 519.}}
\def\BeBT{\rrr\BeBT {B.~Gato-Rivera and A.N.~Schellekens,
Commun.~Math.~Phys. \und{145} (1992) 85 }}
\def\ADE{\rrr\ADE {
M.~Kreuzer and H.~Skarke, ~ADE string vacua with discrete torsion, \rB
CERN-TH.6931/93 preprint}}
\def\va{\vec\alpha}
\def\vb{\vec\beta}
\def\vc{\vec\gamma}
\def\aq{a(\vec q)}
\hfuzz= 6pt
\def\Zbf{{\bf Z}}
\def\mod{{\rm ~mod~}}

\def\half{\coeff12}

\def\GCD{{\rm GCD}}

\def\X{{\cal X}}
\def\Mult{{\rm~Mult}}
\pubnum={{93-13}}
\date{May 1993}
\pubtype{CRAP}
\line{\hfill CERN-TH.6912/93}
\vskip -1.truecm
\titlepage
\message{TITLE}
\title{\fourteenbf
Simple currents versus orbifolds with discrete torsion --
     a complete classification
}
\author{M. Kreuzer}
\CERN
\andauthor{A. N. Schellekens}
\line{\hfil \it NIKHEF-H, P.O. Box 41882, 1009 DB Amsterdam,
The Netherlands  \hfil}
\abstract {
  We give a complete classification of all simple current modular invariants,
  extending previous results for $(\Zbf_p)^k$ to arbitrary centers. We obtain
  a simple explicit formula for the most general case. Using orbifold
  techniques to this end, we find a one-to-one correspondence between simple
  current invariants and subgroups of the center with discrete torsions.
  As a by-product, we prove the conjectured monodromy independence of the
  total number of such invariants.
  The orbifold approach works in a straightforward way for symmetries
  of odd order, but some modifications are required to deal with
  symmetries of even order. With these modifications the orbifold
  construction with discrete torsion is complete within the class
  of simple current invariants. Surprisingly, there are cases where
  discrete torsion is a necessity rather than a possibility.}

\vskip 1.5 truecm
\vfill
\line{CERN-TH.6912/93\hfill} \vskip -3pt
\line{NIKHEF--H/93-13\hfill} \vskip -3pt
\line{June 1993\hfill}
\endpage

\chapter{Introduction}

The problem of classifying and enumerating all modular invariant
partition functions of a given conformal field theory has been
studied intensively during the last five years, but is still far
from solved. However, there is one subclass of invariants that is
almost under control, namely the simple current invariants. Simple currents
\ScYa\ correspond to primary fields that upon fusion with any other field yield
just one field. It is easy to see that the presence of simple
currents implies that the conformal field theory has an abelian
discrete symmetry called the center. A modular invariant partition
function is called a simple current invariant if all fields that are
paired non-diagonally are related by simple currents.

Although not all modular invariants are of this type, experience suggests
that exceptions are rare. Hence by enumerating all of them one has
probably listed most of the possible invariants of a given conformal
field theory. The total number of such invariants grows very rapidly
with the number of simple abelian factors in the center, a
situation that is typical for tensor products of basic building blocks.
Investigations of large classes of modular invariants that can be
obtained with simple currents have been presented for example in
\multref\FKSS{\ScYd\ADE}.

The systematic study of simple current invariants was only partly
complete up to now. In \BeBe\ all the pure automorphisms have been
classified for any center. In \BeBT\ all invariants have been
classified for centers of the form $(\Zbf_p)^k$ (and products thereof),
where $p$ is a prime. What is still missing is a complete classification
for centers containing factors $\Zbf_{p^n}$.

Completing this classification is one of the goals of this paper.
Another goal is to find an explanation for a phenomenon that was hard
to understand from the point of view of \BeBT. In that paper the
modular invariants were constructed by (a) classifying all possible
extensions of the chiral algebra, (b) determining the allowed
'heterotic' combinations of different algebras and (c)
superimposing all allowed automorphisms determined in \BeBe. Then the
total number of invariants was calculated by adding up all these
different kinds of solutions, a rather laborious computation.
It turned out that the total number of invariants is only a
function of the group structure of the center, and does not depend
on the spins and relative monodromies of the currents.

To formulate the latter statement more precisely, consider a
center generated by currents $J_i, i=1, \ldots, k$. The
monodromies of these currents define a symmetric matrix
$$ \tilde R_{ij} = Q_i(J_j) = Q_{j}(J_i) \ , $$
where $Q_i(a)$ is the monodromy phase of the
current $J_i$ with respect to a field labelled $a$, which we will call
the charge of $a$ (for more details
we refer to \ScYa\ or the review \ScYg).
The matrix elements of
$\tilde R$ are quantized as
$$   \tilde R_{ij} ={ \tilde r_{ij} \over N_i },\ \ \ \
\tilde r_{ij} \in \Zbf \ . $$
Note that $\tilde r_{ij}$ is defined modulo $N_i$.
Furthermore $\tilde r_{ij}$  must be
quantized in units of $ N_i / \GCD(N_i,N_j)$ because of the
symmetry of $\tilde R$.  This is equivalent to $\tilde r_{ij} N_j = 0 \mod
N_i$.
The total number of invariants was found to be independent of $\tilde R$,
provided one
reduces the center to a subgroup, the `effective center'.
This is obtained  by removing all
currents whose spin, multiplied by the order of the current, is not
an integer. This eliminates for example the simple currents of
$SU(2)$ at odd levels. In general it eliminates all currents that cannot
preserve $T$-invariance, and hence cannot play a r\^ole in
constructing modular invariants.
The conformal weight of
a current combination $J_1^{\alpha_1} \ldots J_k^{\alpha_k}$
(henceforth denoted $[\vec \alpha]$) can be expressed in terms of
a slight generalization of the matrix $\tilde R$:
$$ h([\vec \alpha]) =  \half\sum_{ij} \alpha_i R_{ij} \alpha_j +
                           \half \sum_i r_{ii} \alpha_i\ , \mod 1 \eqn\Cwt
  $$
where $\tilde R = R \mod 1$. Note that the conformal weight changes by
$\half \alpha_i (\alpha_i + N_i)$ if we change $R_{ii}$ by 1. If $N_i$
is even this may be a change by a half-integer, which is not equivalent.
Thus $R$ contains more information
than $\tilde R$, and its diagonal matrix elements
are defined modulo 2, not 1.
A necessary and sufficient condition for
an effective center is that the diagonal elements $r_{ii}$
are even (note that if $N_i$ is odd all matrix elements
$r_{ij}$ can be chosen even since they are defined modulo $N_i$).
Henceforth the word 'center' always means
'effective center'.

For effective centers $(\Zbf_p)^k$, $p$ prime, it was found that,
even though the separate numbers of different kinds of
invariants (see (a), (b) and (c) above) depend strongly on $R$,
the total number does not, and is given by
the simple formula
$$  N_{\rm tot} = \prod_{i=0}^{k-1} (1 + p^i) \ . $$
Empirically, this phenomenon appears to hold also for more general
centers (\ie\ with $\Zbf_{p^n}$ factors), but there are two
difficulties with pursuing this further. The first is that
although points (a) and (c) have already been solved in general,
point (b) has not, \ie\
no general rule is known for the allowed heterotic combinations
of different algebras (a necessary condition is that they must have
the same size).
The second difficulty is that in the process of
adding up all invariants for all possible monodromies more and more
different cases have to be considered separately. Clearly  a
simpler approach is needed.

Such an approach is already available for $\Zbf_N$, for any $N$.
In this case the classification of \BeBT\ still applies, since
a subgroup is uniquely defined by its size, so that there cannot be
any heterotic invariants. One finds that
the number of invariants is in one-to-one correspondence with the
subgroups of the center. Furthermore, a universal formula exists that
gives all possible invariants \ScYb. To write down this formula one
specifies a subgroup
${\cal H}$ of the center, which is generated by a current $J$. Then the
non-zero values of $M_{ab}$,
the multiplicity of the module
$\ket{a}\otimes\ket{b}$, are given by\foot{This formula is valid
provided that $r$ is even, as discussed above, and that $Q$
is defined modulo 2 in terms of $R$.}
$$ M_{a,J^na} =
    \Mult(a)\ \delta^1(Q(a) + {1\over2} n Q(J) ) \ .\eqn\ZNformula $$
Here $a$ labels a primary field, $J^n a$ is the field $a$ acted upon
$n$ times by the current $J$, and  $Q$ is the charge with respect to $J$.
If $J$ acts without fixed points on $a$ $\Mult(a)=1$, and otherwise
$\Mult(a)$ is equal to the number of copies of $a$ that one
encounters on a standard-length orbit. This formula does depend on
the monodromies via $Q(J)$, but the {\it number} of solutions clearly does
not.

Formula \ZNformula\ was originally obtained by applying orbifold
twists to the discrete symmetries of the center, and hence it is
natural to look in that direction for a more general formula.
Orbifolds are not limited to $\Zbf_N$ groups, and can in fact be
written down for any subgroup of the center. This is still not
enough because the number
of invariants is in general larger than the number of subgroups.
However, as Vafa \Vaf\ has shown, there are more general orbifolds one
can write down because one can allow
phases known as 'discrete torsion'\rlap.\foot{Historically the name refers to
discrete values of a $B_{ij}$ background field. Such an interpretation
is not always available in a straightforward way in arbitrary
conformal field theories, but it is natural to use the same name
in general.}

It is instructive to count the number of orbifold invariants including
discrete torsion. Suppose ${\cal N}_a(k)$ is the number of $(\Zbf_p)^a$
subgroups in $(\Zbf_p)^k$. This quantity satisfies the
recursion relation ${\cal N}_a(k+1)=p^a {\cal N}_a(k) + {\cal N}_{a-1}(k)$.
For each subgroup with $a$ generators the number of discrete torsion
coefficients according to \Vaf\ is equal to $p^{a(a+1)/2}$. Hence the
total number of invariants is
$$ \sum_a {\cal N}_a(k) p^{a(a+1) \over 2 } = \prod_{i=0}^{k-1}(1+p^i) \ . $$
This is precisely the result of \BeBT. Since the latter was shown to
be complete, and since the orbifold invariants are all different,
this establishes a one-to-one relation between the two approaches for
centers $(\Zbf_p)^k$.

If the aforementioned monodromy-independence holds, we can already
go much further by considering the case of trivial monodromy (all
currents are mutually local and have integral spin). In that case
all subgroups ${\cal H}$
of the center ${\cal C}$
can occur as extensions of the algebra. We denote the total number of
subgroups isomorphic to ${\cal H}$ in ${\cal C}$ as
${\cal {\cal N}}({\cal H},{\cal C})$.
Once the algebra is extended, the quotient group ${\cal C} / {\cal H}$
survives as the
center of the new theory. The  number of pure automorphism of a theory
with trivial monodromy, center ${\cal H}$, and generated by
currents of order $N_i$
is given by \Scha
$$ A({\cal H}) = \prod_{i < j} {\rm~GCD}({N_i, N_j}) \ .\eqn\AutCount $$
If all currents are mutually local, the left and right algebras must
be the same, since there is no way to project anything out.
The total number of invariants is thus equal to
$$ \sum_{\cal H} {\cal N}({\cal H},{\cal C}) A({\cal C}/{\cal H}) \ . $$
To get the number of orbifold invariants one also considers all
possible subgroups, but then one counts the number of allowed torsions
{\it within} each group. This number is again given by \AutCount.
Hence in this case we get a total which is equal to
$$ \sum_{\cal H} {\cal N}({\cal H},{\cal C}) A({\cal H}) \ . $$
Although this looks different, obviously
${\cal N}({\cal H},{\cal C}) = {\cal N}({\cal C}/{\cal H}, {\cal C}) \ , $
so that the result is the same. Once again this establishes a one-to-one
correspondence.

Although roughly correct, there
are still some serious flaws in the foregoing arguments.
First of all, it is not straightforward to define an orbifold (without
torsion) for arbitrary subgroups of the center. Although the center
defines a symmetry of the theory, not every symmetry defines an orbifold.
Furthermore the center is {\it a priori} a symmetry of a chiral half
of the theory. It is not completely trivial to find a related symmetry
that acts on the complete theory, and satisfies level matching \Vaf.
It turns out that currents of even order are especially hard to deal
with. In some cases, an orbifold description in the sense of
\Vaf\ does not seem to exist, although we can write down something
similar. There are also cases where one cannot really define an orbifold
without torsion, and where torsion is needed to write down any
non-trivial invariant.
Secondly, since the simple current classification for $\Zbf_{p^n}$ is
not complete for arbitrary monodromies,
and the monodromy independence only conjectured, there
remains a possibility that something is overlooked.

We will overcome the second point by proving directly that
orbifolds with discrete torsion produce all possible simple
current invariants. This requires some refinements of the arguments
of \Vaf, that were not intended to be a proof of completeness, but
only of existence, and that in addition were presented for theories built
out of free bosons or fermions, which is not the case for us. Of course
the issue of completeness cannot
be addressed in a practical manner if one studies orbifolds of tori, since
one moves between different conformal field theories by the introduction
of twist fields. In our case, however, we are dealing with a fixed
set of primary fields from which all the invariants are constructed.

The final result is an extremely simple formula that generalizes
\ZNformula, and yields all simple current invariants in all cases.

In the next section we will present the completeness proof using
orbifold techniques. In section 3 we analyse the resulting invariants
to find extensions of the algebra or pure automorphism, and we will
make the relation with the results of \BeBe\ and \BeBT\ more precise.

\chapter{Orbifolds}

The main task in this section is to find a translation of \Vaf\ that
generalizes
it beyond the free boson/fermion theories for which it was written.
To illustrate this we start with the simplest case, a
theory with single orbit $\Zbf_N$, $N$ odd but
not necessarily prime. Suppose there exists a non-diagonal modular
invariant. If it is a simple current invariant, any non-diagonal
term must be of the form
$$ \X_a \X_{La}^*\ , $$
where $L$ is some simple current. If $N$ is odd, $L$ can be written
as a square of a simple current $K = L^{{N+1 \over 2}}$. Furthermore the
field $a$ can be written as $K^c b$ for some definite $b$. Hence the
off-diagonal term takes the form
$$ \X_{K^c b} \X_{K b}^*\ . $$
Inspired by this and
by \Vaf, we consider now the following set of "twisted'' partition
functions
$$ P[0,n](\tau) = \sum_a \X_{(J^c)^n a}(\tau) \X_{J^n a}^*(\bar\tau) \ , $$
where the sum is over all fields. These partition functions correspond
to $(0,h)$ in the notation of \Vaf. To find the analog of the full $(g,h)$
is now simply a matter of making modular transformations. For example
$$ P[0,n](-{1\over\tau}) =
  \sum_{a,b,c} S_{(J^c)^n a,b} S_{J^n a,c}^* \X_b(\tau) \X_c^*(\bar\tau)\ , $$
using the modular transformation properties of the characters. Now we
use the formula for $S$ from \ScYb\ and \Alig:
$$ S_{J^n a,b} = e^{2\pi i n Q(b)} S_{ab}\ . $$
Furthermore we use $Q_c = - Q $ to get
$$ P[0,n](-{1\over\tau}) =
   \sum_a e^{-2\pi i n (2 Q(a)) } \X_a(\tau) \X_a^*(\bar \tau)$$
The result should be equal to $P[-n,0](\tau)$.
The same computation can be done for the transformation $\tau \rightarrow
\tau + 1$. The results of combining these transformations can be
summarized by defining
$$P[m,n](\tau) = \sum_a e^{2 \pi i (2 m Q(a))}
 \X_{(J^c)^n a}(\tau) \X_{J^n a}^*(\bar\tau) \ , $$
It is easy to verify that this transforms as
$$P[m,n]({a\tau + b \over c\tau+ d})=
  P[am+bn,cm+dn](\tau) \ . $$
Note that $2 Q(a)$ is
simply the charge of a field
$\Phi_{(J^c)^n  a}(z) \Phi_{(J^n a)}(z^*)$ with respect to
the current $J J^c$, and that this does not depend on $n$:
$$ 2 Q(a) = Q((J^c)^n a) - Q_c (J^n a ) \ . $$
This is a consequence of the fact that $J J^c$ is an integer spin
current. This remark will become relevant in a moment.
Obviously, by summing $P[m,n]$ over a modular orbit one gets
a modular invariant partition function. This argument is exactly
the one used when simple current invariants were originally constructed in
\ScYa, where the operator $J J^c$ was used as a
twist operator. Here the formalism of \Vaf\ can be taken over literally.

Generalizing this to $\Zbf_{N_1} \times \ldots \times \Zbf_{N_k}$
(still with all $N_i$ odd) is
essentially straightforward. We just replace $m$ and $n$ by vectors
$\va$ and $\vb$. Then we define
$$P[\va, \vb] =  \sum_a
e^{4\pi i \va \cdot \vec Q(a) }
\X_{[-\vb]a} \X_{[\vb]a}^* \ , $$
where $[\va]a$ denotes the field obtained by acting with
$J_1^{\alpha_1} \ldots  J_k^{\alpha_k}$ on the field $a$. These functions
transform under modular transformations in the obvious way,
$$  P[\va, \vb]({a\tau + b \over c\tau+ d})=
    P[a\va+b\vb,c\va+d\vb](\tau) \ .      $$

The most general partition
function that we can write down using these functions is
$$ P_C  = \sum_{\va,\vb}  C(\va,\vb) P[\va,\vb] \ ,\eqn\Ansatz $$
where the $C's$ are arbitrary complex numbers.
Here the sum is over all values of $\va$ and $\vb$ covered by the modular
group,
modulo equivalences, \ie\ precisely over all group elements of the center.

The crucial question is: can we write the most general
simple current partition function
in this way, or in other words, are the functions $P[\va,\vb]$
a complete basis for the space of simple current partition
functions\rlap.\foot{By this we mean the set of functions
$M_{ab} \X_a \X_b^*$, with $M_{ab}$ a set of positive integers that
vanish of $a$ and $b$ are not connected by simple currents. Note that
we are not (yet) requiring modular invariance here.}It
is clear that they are not, because each function $P$ contains a sum over
all fields, and hence over all orbits. Therefore we cannot modify
the behavior of
individual orbits by changing the coefficients. However, it is not hard
to see that we {\it can} get any simple current partition function that
satisfies the additional requirement that the off-diagonal elements for
each field $a$ depend only on the charge of $a$.
Fortunately it was shown in \BeBe\ and \BeBT\ that
any simple current partition
function that is modular invariant must have that property,
provided the matrix $S$ is reasonably well behaved
(not too many unexpected zeroes and not too many fixed points fields in
comparison to normal fields).
This can be expressed in terms of a
few
regularity conditions, which in practice are usually satisfied.
Since we have nothing new to say about this we will not dwell on this
point, and refer the reader to \BeBT\ for a more detailed discussion
and some examples of pathologies.
Apart from such pathologies, we
can show that by introducing the coefficients $C$ we are able to get
all possible simple current partition functions
that have a chance of being modular invariant.

To see this explicitly, note that
the sum over $\va$ in
\Ansatz\ is simply a Fourier transformation of the set of coefficients.
Indeed, consider the most general candidate invariant
$$ \sum_a \sum_{\vb} M(a,\vb) \X_{[-\vb]a} \X_{[\vb]a}^*\ .  $$
Now suppose that $M$ depends on $a$ only
via the charge $\vec Q(a) \equiv \vec q$.
This allows us to rewrite the
previous expression as
$$\sum_{\vec q} \sum_{a, \vec Q(a) = \vec q}
\ M(\vec q,\vb) \X_{[-\vb]a} \X_{[\vb]a}^*  \ . $$
Using a Fourier transform in the periodic variable $\vec q$ we can
write\foot{Note
that although one would normally put $2\pi$ rather than $4\pi$ in the
exponent, this makes no difference as long as the orders $N_i$ are odd.
It merely reorganizes the sum.}
$$ M(\vec q,\vb)= \sum_{\va} e^{4\pi i \va \cdot \vec q }
C(\va,\vb) , $$
so that the partition takes the form
$$\sum_{\vec q} \sum_{a, \vec Q (a) = \vec q}
\sum_{\va} C(\va,\vb) e^{4\pi i \va \cdot \vec q}
\X_{[-\vb]a} \X_{[\vb]a}^*\ ,
 $$
which indeed is precisely \Ansatz.

To determine the complex numbers $C$ we follow \Vaf.
In this paper only phases were considered,
but that makes little difference. To determine the coefficients
higher loop modular invariance and factorization
was used in \Vaf. On the other hand,
the work in \BeBe\  and \BeBT\ uses one-loop modular invariance and
positivity.
Of course all these conditions are necessary, and even
though positivity is not imposed as a condition, we will be able
to verify it afterwards. We return to this point at the end
of this section.
The higher genus generalization of $C(\va, \vb)$ is
denoted $C(\va_1,\vb_1; \va_2, \vb_2; \ldots)$, exactly as in \Vaf\ except
that we use an additive notation.

One loop modular
invariance implies
$$ C(a\va+b\vb,c\va+d\vb)=C(\va,\vb)\ . $$
Factorization requires
$$ C(\va_1,\vb_1 ; \va_2,\vb_2) = C(\va_1,\vb_1) C(\va_2,\vb_2)\ . $$
The Dehn twist around a curve connecting adjacent handles yields
$$C(\va_1,\vb_1; \va_2,\vb_2)=C(\va_1+\vb_2-\vb_1,\vb_1;
\va_2+\vb_1-\vb_2,\vb_2)\ , $$
and finally we can normalize $C(0,0)=1$. {}From these conditions one
derives
$$ C(\va_1,\vb_1)C(\va_2,\vb_2)=C(\va_1+\vb_2,\vb_1)C(\va_2+\vb_1,\vb_2) \ . $$
Setting $\va_1 = \va_2 = \vb_2 = 0$ we get
$$ C(0,\vb_1) C(0,0) = C(0,\vb_1) C(\vb_1,0)\ . $$
We would like to conclude from this that $C(\vb_1,0)  = C(0,0) = 1$,
but of course this
is only true if $C(0,\vb_1)\not=0$. This possibility is rejected in
\Vaf\ because all $C$'s are assumed to be phases, but in fact there
are perfectly valid solutions in which some coefficients vanish.
It is not hard to see that these coefficients vanish in such a way
that the non-vanishing ones span only a subgroup of the
center. Of course this is exactly as it should be.
If indeed by introducing $C$'s one can get all possible
modular invariants, one must in particular be able to get all the
invariants corresponding to subgroups, including the diagonal one.
{}From now on we will assume that the $C$'s don't vanish.

The rest of the argument proceeds as in \Vaf.
We find
$$ C(\vb+\va,\vc)=C(\va,\vc)C(\vb,\vc) \ , $$
which implies
$$ C(\va, \vc)^N  = C(N\va,\vc) = C(0,\vc) = C(-\vc,0) = 1 \ , $$
if $N$ is the order of $\va$. This
shows that $C$ must be a phase after all, and is in fact an $N^{\rm th}$
root of unity. The classification of all allowed phase choices is then
exactly as in \Vaf, and will be explained below.
This concludes the proof of completeness for odd orders.

Now we still have to deal with even $N$.
Let us again start with a single factor. In
$\Zbf_N$, $N$ even one cannot (in general) 'take the square root' of
$J$. If $J$ has integer spin this is still not a problem, because one
can work with the 'chirally' twisted sectors  $\X_a \X_{Ja}^*$, and everything
goes through just as before. What happens if $J$ has some other spin?
It seems natural to define
$$P[m,n]= \sum_a e^{2\pi i m Q(a)} \X_a \X_{J^n a}^*\eqn\RightInv$$
Note that choosing $a$ or $J^n a$ as the argument of $Q$ yields a different
result (except if $J$ has half-integer spin).
However, the real problem is that $P$ transforms with a phase that
depends on $n,m$ and $a,b,c,d$ in a very complicated, way. In fact, we
do not even know a
general formula for that phase,
except when the current has half-integer spin. Then we find
$$P[m,n]({a\tau + b \over c\tau+ d})=(-1)^{am+bn+m}(-1)^{cm+dn+n}
  P[am+bn,cm+dn](\tau) \ . $$
Keeping theses phases in a sum over a modular orbit, we find for example
if $N=2$ that
$$ P[0,0] \pm  ( P[1,0] + P[0,1] - P[1,1] )  $$
is modular invariant. To keep the identity the $\pm$ sign must be taken
to be $+$, and this choice gives the expected automorphism invariant
generated by a spin-${1\over 2}$ current.
Formally this is analogous to the prescription given in \Vaf\
for orbifolding free
fermions, but
half-integer spin simple currents occur in many CFT's that have nothing
to do with free fermions.

Beyond this example the procedure of \Vaf\ becomes really inadequate for our
purpose. For example, consider $SU(4)$ level 2. This has four simple
currents of spin 0, twice ${3\over4}$  and 1. The spin-1 invariant is
easy to get, but there is an extra automorphism generated by the
spin-${3\over4}$ current $J$. One might think that this can be gotten
by using the operator $J J^c$, but that is not true. This just gives
the same spin-1 invariant. The 'correct' answer turns out to be
to modify \RightInv\ by using in the exponential not
$Q(a)$ nor $Q(J^na)$ but the average:
$$P[m,n]= \sum_a e^{2\pi i m {1\over2}(Q(a) + Q(J^n a))} \X_a\X_{J^n a}^*\ .$$
This expression is ill-defined as it stands, since charges are defined
modulo 1. By writing the exponent as $i\pi (2 Q(a) + Q(J^n))$ we see that
in particular the second term requires more care. It turns out that
under $S$ the transformation is as for odd $N$, for any valid definition of
$Q$.
However, the $T$-transformation is sensitive to the precise definition
of $Q(J)$. If we define $Q(J)$ with the same matrix $R$ used also in
the definition for the conformal weight (see \Cwt) one gets the
following transformation under $T$:
$$P[m,n](\tau+1) = e^{2\pi i{nr\over 2}} P[m+n,n](\tau) \ .\eqn\AvInv $$
The extra phase disappears if $r$ is even, which indeed is a necessary
condition for the existence of the modular invariant, and is automatic
for the effective center.
Without the phase,
$P[m,n]$ transforms exactly as before, and by summing over a modular
orbit one gets the desired invariant,
which is precisely \ZNformula.

Incidentally, one may ask what happens if, for $r$ odd, one simply keeps
the phase and sums over the modular transformations. Then the
sum will have to extend
over twice the modular domain, because $P[m,n] \not = P[m,n+N]$ for
odd $m$. The result is that the terms with improper periodicity cancel
out, and one finally gets some invariant corresponding to a smaller
subgroup.

What is somewhat disturbing about \AvInv\ is that the phase factor
does not seem to correspond to any symmetry operator $g$ acting on
the states of the theory, like the `$g$' of the orbifold approach. An
operator that produces such eigenvalues would be
$\sqrt{J J^c}$, which looks rather unpleasant. Furthermore the
'twisted' sectors are not created with this operator, but with $J$
acting on the right-moving sector.
Nevertheless the proof of modular invariance at one loop is
completely rigorous. Note that these arguments, as well as the following
ones, hold for odd as well as even orders.

Now
let us consider the general case. Clearly we would like to define objects like
$$P[ \va, \vb] =  \sum_a
e^{\pi i \va \cdot (\vec Q(a) + \vec Q([\vb]a))  }
\X_{a} \X_{[\vb]a}^* \ . $$
It is easy to show that these functions transform correctly under $S$, and
under $T$ they transform with an extra phase
$\exp(\pi i \sum_i \beta_i r_{ii})$, that is equal to 1 if $r_{ii}$ is
even, which is always true for the effective center. Here and in the
following all charges of currents are defined using the matrix $R$ of
\Cwt, whose diagonal elements are defined modulo 2. As before, this
is imposed by $T$-invariance. One loop modular
transformations invariance do not impose any particular choice for the
$\mod 2$ ambiguity in the off-diagonal elements of $R$, as long as $R$ is
symmetric modulo 2. Note that the conformal weights also do not fix
this ambiguity. We simply make an arbitrary choice to fix this ambiguity
in $R$ (which amounts to a sign-ambiguity for $P$). For odd $N_i$ it
is convenient to choose all $r_{ij}$ even.

First we will show, as before,
that any modular invariant partition function can
be written in this basis.
Any simple current invariant has the form
$$ \sum_{a,\vb} M(a,\vb) \X_a \X_{[\vb] a}^* \ . $$
As before we use the result of \BeBT\ that
$M$ depends on $a$ only via
the charge $\vec Q$ of $a$. Then we
can write this as
$$ \sum_{\vec q} \sum_{a(\vec q)} M(\vec q,\vb) \X_{\aq} \X_{[\vb]\aq}^*\ . $$
Now we Fourier-transform $M$ with respect to $\vec q$:
$$ M(\vec q,\vb) = \sum_{\va} \tilde C(\va,\vb) e^{2\pi i \va\cdot\vec q}\ .$$
This will not yield exactly the functions $P(\va,\vb)$ we are trying
to get, so what we have to do is redefine the coefficients as follows
$$ \tilde C(\va,\vb) = e^{\pi i \va\cdot R \cdot \vb}  C(\va,\vb) \ . $$
Substituting this we have indeed written the modular invariant as
$$ \sum_{\va,\vb}  C(\va,\vb) P[\va,\vb] \ .\eqn\Decomp $$
This shows that we can indeed write every modular invariant in this basis.
Note that the  ambiguities in the definition of the off-diagonal elements of
$R$ modulo 2 cancel between $C$ and $P$.

Now we have to answer the question: for which choices of $C$ do we get a
modular invariant. First of all, suppose one knows one valid choice. Then
one can factor it out of all the $C$'s, and the remainder should then
satisfy the same equations as the discrete torsions introduced earlier.
Now previously there
was always a trivial solution, namely $C=1$. This may appear to be true
here as well, but on closer inspection it is not.

Eq. \Decomp\ is summed over a set of vectors belonging to a domain that
covers all inequivalent vectors exactly once. It is manifestly
modular invariant under all transformations that map these vectors
within this domain. If $P$ has the proper periodicity it is then
modular invariant. However, $P$ does not always have the right
periodicity. The factor that may violate the periodicity is
$$ e^{i\pi  \va \cdot \vec Q(\vb) } = e^{i \pi \va \cdot R \cdot \vb}\ . $$
In general, $R_{ij} = {r_{ij} \over N_i} $. Thus if we shift $\va$
by a period $N_k$ for some $k$ we get a phase $\exp(i\pi r_{kl} \beta_l)$.
If all the matrix elements of $r$ are equal to an even integer this
phase is equal to 1. This property of $r$ can always be arranged to hold for
odd $N_i$. Furthermore it always holds for the diagonal elements if we restrict
ourselves to the effective center, but it need
not hold for the off-diagonal ones.

Clearly the choice $C=1$ will not do in that case. We need some choice
of $C_0$ that respects modular invariance within the
domain described above, but also has the wrong periodicity to compensate
the wrong periodicity of $P$.
It doesn't matter how we get such a $C_0$, and it also doesn't matter
which choice we make. If someone else makes a different choice, it
follows from the above arguments
that his choice can be obtained from ours be multiplication
with a discrete torsion factor. Since $P$ transforms 'correctly'
within the domain, $C_0$ must transform like a discrete torsion
factor within the domain.
An obvious choice for $C_0$ is
$$ C_0[\va,\vb]=e^{i \pi \va \cdot E \cdot \vb} \ ,\eqn\Czero $$
where $E$ is an antisymmetric matrix satisfying $E_{ij}=R_{ij}$
for $j > i$. This does indeed satisfy all conditions of a
discrete torsion, and has the wrong periodicity whenever $R$ does.

Apart from this subtlety, the rest of the argument goes exactly as for
odd order. In both cases,
the general discrete torsion can be written as
$$ C(\va,\vb) = e^{2 i \pi \va \cdot e \cdot \vb} \ , $$
where $e$ is an antisymmetric matrix with matrix elements
$$ e_{ij}= {\epsilon_{ij}\over N_i} = - {\epsilon_{ji}\over N_j} \ \ \ ,
{\rm ~with}\ \epsilon_{ij} \in \Zbf \ . \eqn\equant $$
That this is the most general form follows from \Vaf, and can be
seen easily by defining
$C= e^{\Gamma}$. Then the equations for $C$ yield
$$ \Gamma(\va + \vb, \vc) =  \Gamma(\va,\vc) + \Gamma(\vb,\vc) $$
and
$$ \Gamma(\va,\vb) = - \Gamma(\vb,\va)\ ,  $$
so that $\Gamma$ is linear in both its arguments. Furthermore
$$ \Gamma(\va,\va) = 0 \ .$$
Thus $\Gamma$ is a bilinear antisymmetric object, and must be of
the form $\va \cdot e \cdot \vb$ for some $e$. The matrix elements
of $e$ are then restricted
to the form \equant\ by requiring the correct periodicity.

Now we can absorb $C$ into $P$
and write the most general
invariant as follows
$$\sum_a \sum_{\va,\vb} e^{2\pi i \va \cdot \vec Q(a)}
                  e^{2 \pi i \va \cdot X \cdot \vb}
                  \X_a \X_{[\vb] a}^* \ ,\eqn\GenInv $$
where $X$ is a matrix satisfying $X+X^T= R$\rlap,\foot{Note that
$X$ is defined modulo 1 and that $R$ is defined modulo 2 on the
diagonal and modulo 1 elsewhere. The equations are defined with exactly
the same periodicities as $R$. In the following these
periodicities will be omitted from the equations.}
$$ X_{ij}= {\chi_{ij} \over N_i }, \ \ \ \ \ \ \chi_{ij} \in \Zbf\ \ .
\eqn\Xquantization $$
Note that the matrix elements of $X$ are quantized precisely like those
of $\tilde R$: since $X$ is defined modulo integers, $\chi_{ij}$ must be
defined modulo $N_i$; furthermore, in order to satisfy $X+X^T=R$,
$\chi_{ij}$ must be proportional to $N_i / \GCD(N_i,N_j)$, and hence
$\chi_{ij} N_j = 0 \mod N_i$.
In our previous notation
$ X = e + {1\over 2} R $ or $e+{1\over2} ( R+E)$, where
the second expressions is used if $R$ is not divisible by 2. Note
that the r\^ole of the discrete torsions is simply to provide
an antisymmetric part to $R$.
The invariant \GenInv\ is not yet normalized, but it is easy to see
that one must simply divide by the order of the group.

This result can be simplified further by observing that
the sum over $\va$ is just a $\delta$-function. This yields the final,
and undoubtedly most simple and elegant formula for the modular
invariant partition function. Before presenting it, let us
summarize the main result of this paper.

Suppose one has a conformal field theory with simple currents
generating a center ${\cal C}$.
Then the complete set of simple current invariants
of that theory can be obtained by the following procedure
\pointbegin Choose any subgroup ${\cal H}$ of ${\cal C}$.
\point Choose a basis of currents $J_1, \ldots, J_k$ that generate ${\cal H}$.
\point Compute the current-current monodromies $R_{ij}$ in that basis.
\point Choose any properly quantized matrix $X$ (see \Xquantization) whose
symmetric part is ${1\over2}R \mod 1$ (in other words $X+X^T = R $).
The modular invariant partition function corresponding to this choice
is then given by a matrix whose only non-zero elements are
$$ M_{a,[\vb]a} = \Mult(a) \prod_i \delta^1 ( Q_i (a) + X_{ij} \beta_j )\ ,
\eqn\DeltaForm $$
where $\delta^1$ is equal to 1 if its argument is an integer, and
vanishes otherwise. The factor $\Mult(a)$ appears because $a$ may be
a fixed point of some currents.
In that case the $\beta$-sum in \GenInv\ includes all
terms involving $a$ more than once, and $\Mult(a)$ is the number of times
this happens.  This is the generalization of \ZNformula\ to more than one
factor.

As we mentioned earlier, this is the complete set of solutions to a
different set of conditions than those considered in \BeBT\ and \BeBe\ (and
in most other papers on modular invariance of conformal field theories
other than those built out of free bosons an fermions). Usually one
tries to determine all positive and properly normalized matrices $M$
that commute with $S$ and $T$, a given set of representations of the
(one-loop) modular group. In our case it turned out to be convenient
to replace the positivity condition by a higher genus condition. This goes
a little bit against the original spirit, since it requires more
information than just $S$ and $T$. Indeed, we do not even know how
to write down explicitly
the complete higher genus modular invariance conditions
for a generic CFT of which {\it a priori} we only know $S$ and $T$.
In the special case of simple current invariants the orbifold
analogy strongly suggested a higher genus transformation rule which
turned out to give a very effective shortcut in the completeness proof
(an interesting question is whether a similar higher genus approach could be
applied successfully to the long-standing classification problem of
exceptional invariants). Since the resulting invariants are
manifestly positive, they form at least a subset of the solutions to
the original  genus-1 problem. Since in all known cases we find
in fact {\it all} solutions, it is natural to
conjecture that this will be true in general. If this
conjecture turns out to be wrong, the exceptions most likely do not
correspond to well-defined CFT's, unless their higher genus
behavior is unexpectedly subtle.

Finally, note
that for some subgroups ${\cal H}$ discrete torsion is {\it required}
to get any invariant at all.
Consider for example $SO(8)$ level 1. This
has a center $\Zbf_2 \times \Zbf_2$, and the matrix $R$ is
$$ {1\over 2} \pmatrix{ 2 & 1 \cr 1 & 2 \cr} \ , $$
so that indeed the periodicity problem occurs here. There are in total
six invariants, corresponding to the six permutations of the conjugacy
classes $(v), (s)$, and $(c)$. The trivial permutation (the diagonal
invariant) belongs to the identity subgroup. There are three $\Zbf_2$
subgroups, which correspond to the three permutations of order 2. And
finally, there are two invariants corresponding to the $\Zbf_2 \times
\Zbf_2$ subgroup, namely the two cyclic permutations. Clearly these two
are completely equivalent, and there is no sense in which one of them
has discrete torsion, and the other does not. Indeed, the difference
between them is simply that in one case one uses $C_0$, as in
\Czero, and in the other case the complex conjugate. In both cases
there is discrete torsion.

\chapter{Chiral algebras and Automorphisms}

We have now classified all simple
current invariants for any center. In this chapter we will investigate
their properties more closely, to determine the
extensions of the chiral algebra and the automorphisms they imply. This
will also clarify the relation with \BeBe\ and \BeBT\ where the
various kinds of invariants were considered separately.
Finally, we will study products of invariants of the form \DeltaForm\
and explicitly prove closure under multiplication.

\section{Extended chiral algebras}

{}From \DeltaForm\ one can immediately read off the extensions of the
right algebra, \ie\ the nonvanishing matrix elements $M_{0,[\va]}$. The
condition is simply that $X \va = 0 \mod 1$, or in other words that
$\va$ is in the kernel of $X$.
To see how the left algebra is extended we need the following
reflected version of \DeltaForm:
$$ M_{[\vb]a,a} = \Mult(a) \prod_i \delta^1 ( Q_i (a) + X_{ij}^T \beta_j )\ .
\eqn\DeltaFormR $$
Clearly the currents in the left algebra form precisely the kernel
of $X^T$.

If a current $\va$
appears in the left as well as the right algebra, then $X\va=X^T \va=0$,
so that $R\va=0$. This implies that it is local with respect to all
currents in the subgroup ${\cal H}$. Obviously the converse is also
true: if a current is local with respect to all other currents in a
subgroup, it is either in both the left
 and the right
algebra or in neither, in all
invariants belonging to that subgroup (and subgroups thereof).
Hence
if a current is local with respect to all simple currents in the full
center ${\cal C}$ it can not appear 'heterotically' in any simple current
invariant.
This was proved in a different way in \BeBT. Furthermore
it is true that if a current
is non-local with respect to at least one other currents in a subgroup,
it cannot appear simultaneously in the left and right algebras in any of the
invariants corresponding to that subgroup (and all subgroups containing it).

\let\ex=\times \def\al{{\ca_L}} \def\ar{{\ca_R}} \let\ZZ=\Zbf

\def\ca{{\cal A}}  \def\cc{{\cal C}} 
 \def\cf{{\cal F}} \def\cg{{\cal G}} \def\ch{{\cal H}}
\def\ci{{\cal I}}

\let\a=\alpha \let\b=\beta \let\g=\gamma \let\d=\delta 
    
\let\l=\lambda

   \let\D=\Delta

\def\mao#1{\mathop{\rm #1}\nolimits}  \def\Im{\mao{Im}}  \def\mod{\mao{mod}}
\def\({\left(}     \def\){\right)}    
\def\0{\over }     \def\1{\vec }      \def\2{{1\over2}} \def\Ker{\mao{Ker}}

\def\vb{\1\b} \def\vc{\1\g}   

It is of course interesting to see if we can get any general restrictions on
the possible combinations of left and right chiral algebras. We
begin with an instructive example: Consider a center $\cc=\ZZ_9\ex\ZZ_9$
with monodromy matrix  $R_{12}=1/9$, $R_{11}=R_{22}=0$ and its diagonal
subgroup $\ch=\ZZ_9\ex\ZZ_3$. Choosing the discrete torsions such that
$$       X=\left(\matrix{0&0\cr {1\over3} &0\cr}\right)       $$
we have $\al=\ZZ_9$ and $\ar=\ZZ_3\ex\ZZ_3$, so that the chiral algebras
obviously need not be isomorphic. Furthermore, in this case $\cc/\al=\ZZ_9$
and $\cc/\ar=\ZZ_3\ex\ZZ_3$ are not isomorphic either, but $\ch/\al=\ZZ_3$ and
$\ch/\ar=\ZZ_3$ are. The latter isomorphism is in fact a general feature.

To see this note that we are dealing with two kinds of groups: A current
subgroup $\ch$ is a set of (equivalence classes of) currents
that closes under fusion. The corresponding charge subgroup $\ch^*$
is the additive group
of charges modulo 1 that are allowed by $\ch$.
These two groups are of course isomorphic.
The matrix $X$ maps $\ch/\ar$ linearly onto a subgroup of $\ch^*$,
whereby the group structure is obviously preserved.
Now if $q=X\b$ is in the image of $X$, then $\l q=\l X \b=0$ for all
$\l\in\al$. Hence $\mao{Im} X$ is a subgroup of $(\ch/\al)^*$,
and $\ch/\ar$ must
be isomorphic to a subgroup of $\ch/\al$. Replacing $X$ by $X^T$ we also have
the opposite inclusion, and thus the two quotients must be isomorphic.

We can also reverse the argument: Given a subgroup $\ch$ of the
center and two subgroups $\al$, $\ar$ of $\ch$ with isomorphic quotients,
we can choose any isomorphism between $\ch/\ar$ and $(\ch/\al)^*$ to define
a unique matrix $X$ that is appropriately quantized and has $\ar$ and $\al$
as its kernels. In practice, however, this is of little use since one always
starts with a definite monodromy matrix $R$. A particular extension thus
can occur if and only if
$X+X^T$ equals $R$ for at least one of the isomorphisms.

Given two subgroups ${\cal A}_L$ and ${\cal A}_R$
of equal order, the requirement that there is any group
$\ch$ such that $\ch/\al$ and $\ch/\ar$ are isomorphic is, of course,
rather trivial. But as we start
with a given center $\cc$ we get, in general, at least a restriction on
the subgroups that can possibly produce a given heterotic combination.
For some centers we even get more: Consider, for example, $\cc=\ZZ_{p^A}\ex
\ZZ_p\ex\ZZ_p$, where $p$ is a prime number. It is obvious that we can
never get a modular invariant with $\al=\ZZ_{p^2}$, $\ar=\ZZ_p\ex\ZZ_p$
and $\al\cap\ar={1}$, because the smallest group that contains $\al$ and
$\ar$ with isomorphic quotients is $\ZZ_{p^2}\ex\ZZ_{p^2}\ex\ZZ_p$, which is
not a subgroups of $\cc$.

Unfortunately we do not know a simple general rule that covers all
possible left-right combinations of algebras, like the rule
formulated in \BeBT\ for $(\Zbf_p)^k$. It seems that the general
problem is rather complicated. The rules explained above are
necessary, but not sufficient, and are in any case of some help.
We emphasize, however, that although there is no simple rule,
all possibilities can be enumerated easily be generating all matrices $X$.

\section{Automorphisms}

If the kernels of $X$ are trivial, it corresponds to an automorphism. In
\BeBe\ automorphisms were defined by means of
 an integral
matrix $\mu_{ij}$,
defined modulo $N_j$ and satisfying  $N_i \mu_{ij} = 0 \mod N_j$. (Note
that the transpose of $\mu$ has the same quantization as the matrices
$r$ and $\chi$.) Ordinary matrix multiplication closes on such a set
of matrices: if $\mu_1$ and $\mu_2$ satisfy the quantization rule, so
does $\mu_1 \mu_2$. We can define an identity matrix:
$\delta_{ij} = 1 \mod N_j$ if $i=j$ and $\delta_{ij}=0\mod N_j$ if
$i \not = j$ (note that the definition of this element is {\it not}
symmetric in $i$ and $j$). Furthermore we can define an inverse for
a subset of all matrices $\mu$.
The kernel of $\mu$ is defined as the set of vectors $\vb$
satisfying $\sum_i \beta_i \mu_{ij} = 0 \mod N_j$. The matrix $\mu_{ij}$
specifies for each basic charge $i$ by how much the automorphism moves
it in the direction $j$. The kernel corresponds thus to those
charges that are not moved at all. This definition of the kernel is
identical to the one given above for $X$ provided that one writes it in terms
of the matrix ${1\over N} \mu^T$, which  has the same quantization
rule as $X$.

It is easy to show that the left and
right inverse are the same, that an inverse exists if and only if the
kernel is trivial, and that this inverse
is unique modulo the usual periodicities.
The matrix $\mu^T$ belongs to
a similar, but different set on
which the `mirror image' of all these properties
holds. In particular, if $\mu$ is invertible so is $\mu^T$, and
its inverse is the transpose of that of $\mu$.

Not all such matrices $\mu$ define an automorphism, but only those
that satisfy the equation \BeBT
$$ \mu {1\over N} + {1\over N} \mu^T  + \mu R \mu^T = 0 \mod 1 \ . $$
If $\mu$ is invertible we can multiply from the left with
$\mu^{-1}$ and from the right with $\mu^{T-1}$ to get
$$ {1\over N} \mu^{T-1}  + \mu^{-1} {1\over N} + R = 0 \mod 1 \ . $$
Comparing this with the equation for $X$ we see that it is
satisfied by
$$    {1\over N} \mu^{T-1} = - X = - {1\over N} \chi \ ,
 $$
or $\mu = -\chi^{T-1}$.  This can also be derived directly from
\DeltaForm\ and the definition of $\mu$.
Note that ${\chi}$ is
indeed always invertible if it represents an automorphisms.

If $\mu$ were always invertible this would establish a one-to-one
relationship between the two descriptions. However, if $\mu$ is not
invertible, it has a non-trivial kernel, and hence there are
charges on which the automorphism acts trivially. These charges
can always be removed by restricting to a subgroup. By choosing
a small enough subgroup, one can always make $\mu$ invertible.
This reflects a difference in philosophy between the orbifold method
used here and the approach of \BeBe\ and \BeBT: The former works
always within subgroups of the center, whereas the latter constructs
all invariants directly within the full center.
Taking this into
account, we find thus  an exact one-to-one mapping between the
pure automorphisms in both formalisms. Since for automorphisms
both formalisms yield the complete set of solutions to their
respective conditions for any center, this
provides
additional evidence for our conjecture that those conditions are in fact
equivalent.

\section{Products of invariants }

If one multiplies two matrices $M_1$ and $M_2$ that define  a modular
invariant,
the result is obviously modular invariant as well. Furthermore
the product is positive. If $M_1$ and $M_2$ are both simple current
invariants, so is the product. In general, the matrix element $M_{00}$ of
the product matrix is not equal to 1, but at least for simple current
invariants this is always just an overall factor, which can be divided
out. Hence the
complete set of positive simple current invariants must close
under matrix multiplication.

The physical meaning of matrix multiplication in terms of conformal
field theory is not clear in general, and indeed it may well happen
that for exceptional invariants the product does not correspond
to a meaningful CFT
(products of simple current invariants can usually be
interpreted as consecutive orbifold twists). It is also not obvious
that higher loop conditions are automatically satisfied for such a
product, or even how to formulate them.
As we discussed at the end
of section 2, positivity and higher genus invariance are probably
equivalent in the present context. If that is true, our set of invariants
should close under matrix multiplication. Checking closure is in any
case a good test of this conjecture.

This means that there must be an associative product operation for
pairs $(\ch,X_\ch)$, where $R_\ch=X_\ch+X_\ch^T$ has to be the monodromy
matrix for the subgroup $\ch$ of the center. Let us first
consider the simplest case of two invariants $M(\ch,X_\ch)$ and $M(\ch,Y_\ch)$
corresponding to the same subgroup. We then have\foot{For notational
simplicity we assume that $a$ is not a fixed point.}
$$   M(X)_{[\vb]a,[\vc]a}=\d^1(\1Q(a)+X^T\vb+X\vc), $$
where $\d^1(\1q)$ is 1 if all components of the vector $\1q$ are integer.
Thus
$$(M(X)M(Y))_{a,[\vc]a}=\sum_{\vb}\d^1(\1Q(a)+X\vb)~\d^1(\1Q(a)+(R-Y)\vb+Y\vc)
$$
$$~~~~~~~~~~~~~~~~~~~~~~~~ =\sum_{\vb}\d^1(\1Q(a)+X\vb)~\d^1(Y\vc-(X+Y-R)\vb)
\ . \eqn\MXY $$
If the antisymmetric matrix $\D=X+Y-R$ in eq.~\MXY\ is invertible we
can solve for $\vb$ and obtain the product invariant
$M(X*Y)=M(X)M(Y)$ with
$$                      X*Y=X(X+Y-R)^{-1}Y.                       \eqn\XxY $$
There are two reasons why $\D$ may not be  invertible.
First of all there may be vectors $\beta$ that are in the kernel of
$\D$ and also in the kernel of $X$.
Clearly the summand in \MXY\ is
totally independent of such vectors, and performing the sum just
gives an overall factor independent of $\vec Q$ and $\vc$. This factor is
$$N(X,Y) = | \Ker X \cap \Ker \Delta | = |\Ker X \cap \Ker Y^T | =
|\ar(X) \cap \al (Y) |\ . $$
  A trivial example of this situation is to
multiply two identical left-right symmetric pure chiral algebra extensions
($X=Y=R=0$ within $\ch$).

In addition it may happen
that the product $M(X)M(Y)$ cannot be written in the form
$M(Z)$ for any $Z$ with the same subgroup $\ch$, but only for a subgroup
of $\ch$.
This is the case
exactly if the
last $\d$-function in \MXY\ constrains the possible values of
$\1\g$, i.e. if the image of $\D$ does not contain the image of $Y$.
This can be understood as a partial cancellations of the
automorphism actions of $X$ and $Y$. The simplest example is $Y=X^T$,
with both $X$ and $Y$ invertible, so that $X$ and $Y$ define mutually inverse
automorphisms.
Note that the second $\delta$-function implies
a charge independent restriction on $\vc$, which can certainly not be of the
form \DeltaForm.
The reduced subgroup $\ch'$
is thus expected to be the set of vectors $\vc$
for which the equation $Y \vc = \D \vb$ has at least one solution
$\vb$. If there are more solutions they differ by vectors in $\Ker \D$.
We can now perform the sum over
$\vb \in \Ker \D \cap \Ker X$ to
obtain
$$ N(X,Y) \!\! \sum_{\vb \mod \Ker \D} ~~~ \sum_{\vb_0 \in \Ker \D
   \mod \Ker X} \!\!\!\!
    \delta^1(\vec Q(a) + X  \D^{-1}  Y \vc + X \vb_0)
    \delta^1(Y\vc - \D \vb) \ ,\eqn\StepOne $$
where $\D^{-1} Y \vc$
is a formal notation for one representative of the solutions to
$Y \vc = \D \vb$. Of course this set of representatives can be
chosen such that it depends linearly on $\vc$.
The second $\delta$-function restricts $\ch$ to
$\ch'$ for the currents $\vc$.
For a fixed $\vc$, the constraint of the first $\delta$-function
in \StepOne\ has at most one solution $\vb_0$, since we have
already summed over $\Ker \D \cap \Ker X$. The
sum over $\vb_0$ effectively reduces the number of
$\delta$-restrictions. Hence it must be possible to write
the first
$\delta$-function as $\delta^1(P (\vec Q(a) + Z \vc) )$
for some linear map $Z$,  where
$P$ projects to a subgroup $\ch'' = \Im P$.

\def\vnu{{\vec \nu}}
It is easy to see that $\ch'' \supset \ch'$: Consider a vector $\vnu \in \ch'$.
By definition of $\ch'$ there exists some vector $\vb'$ with
$\vnu Y^T=\vb'\D^T$, therefore any vector $\vnu\in\ch'$ is orthogonal to the
term $X\vb_0$ in \StepOne,
$$\vnu \cdot X\vb_0 = \vnu \cdot Y^T \vb_0 = \vb' \cdot \Delta^T \vb_0
                    = - \vb'\cdot \Delta \vb_0 = 0      \ , $$
where we used $\vb_0 \in \Ker \Delta$ and $\D=X-Y^T=-\D^T$.
It follows that the projection
of the $\delta$-function argument on $\ch'$ is consistent with the
sum over $\vb_0$, so that indeed $\ch''$ must contain $\ch'$.

It then only remains to show that the dimensions of $\ch'$ and $\ch''$
are equal.
This can be concluded from the one-loop modular invariance of the product
invariant \StepOne,
as we show in the Appendix, or from a direct computation:
The dimension of $\ch/\ch'$ is given by the number of restrictions on
$\vc$ that come from the second $\d$-function,
which is $|\Im Y| / |Im Y \cap \Im \D|$.
The number $|\ch/\ch''|$ of different charges
that are allowed for a given $\vc$, on the other hand, is equal to
$ |\Ker \D| / |\Ker \D \cap \Ker Y^T| $, \ie\ the number
of vectors $\vb_0 \in \Ker\D\mod\Ker X$.
Now we already know from section 3.1
that  $ \Im Y = (\ch / \Ker Y^T)^* $ for any matrix
$Y$ satisfying the quantization conditions, and therefore also
$ \Im Y \cap \Im \D  =  (\ch / \langle\Ker Y^T,\Ker \D\rangle)^* $,
where $\langle A , B \rangle$ denotes the span of $A$ and $B$.
Altogether we find
$$ {|\ch|\0|\ch'|} = { |\Im Y| \0 |\Im Y \cap \Im \D| } =
   { |\ch| / |\Ker Y^T| \0 |\ch| / |\langle\Ker Y^T,\Ker \D\rangle| } =
   { |\Ker \D| \0 |\Ker Y^T \cap \Ker \D| }={|\ch|\0|\ch''|} .  $$
Hence we have show now that $\ch'=\ch''$, and therefore
$$ M(\ch,X)M(\ch,Y)=|\ar(X)\cap\al(Y)| ~ M(\ch', Z_{\ch'})
      \eqn\OAM $$
for some matrix $Z_{\ch'}$. Roughly speaking, this matrix is nothing
but $X*Y$ defined earlier, but with a choice of representatives for
the inverse map $\D^{-1}$ plus a projection to $\ch'$. Using
once more that the product is in any case one-loop modular invariant
we can
show (see Appendix A)
that $Z$ satisfies $Z+Z^T = R$ in $\ch'$.
The subgroup $\ch'$ can formally be written as
$\ch'=\Im (Y^{-1}\D)\subseteq\ch$, where $Y^{-1}$ is a
one-to-one map from $\Im Y$ to $\ch \mod \Ker Y$ (in general it is thus
not a well-defined map from $\ch$ to $\ch$, but the image of $Y^{-1}\D$ is
nevertheless a well-defined subset of $\ch$).
With a similar
notation,
the extended chiral algebras are given by $\ca'_R=Y^{-1}Y^T\ca_R(X)$, where
of course $\ca_R(Y)\subseteq\ca'_R\subseteq\ch'$,
and by $\ca'_L=(X^T)^{-1}X\ca_L(Y)\supseteq \ca_L(X)$.

The formula \OAM\ for the overall
multiplicity in the product invariant is of course
correct also in the general case of two different effective subgroups $\cf$
and $\cg$. In that case, the effective subgroup $\ch'$ of the product
invariant is obviously a subgroup of the group $\ch$ that is spanned by
$\cf$ and $\cg$. It is thus convenient to refer to a basis of $\ch$, in terms
of which bases of $\cf$, $\cg$ and $\ci=\cf\cap\cg$ can be defined by
matrices $F$, $G$ and $I=I_FF=I_GG$ as explained in Appendix A. Any vector
$\vc \subset \ch$ can be written as
$$\vc=F^T\vc_F + G^T\vc_G\ . $$
Of course if $\cf$ and $\cg$ have a non-trivial overlap
there is more than one way to make this
decomposition.
Now consider a
product $\sum_{\vb}M_{a,[\vb]a} M_{[\vb]a,[\vc]a}$ where the first matrix
is defined in $\cf$ and the second in $\cg$.
The sum over $\vb$ is {\it a priori}
over all of $\ch$, but is restricted to the subset $\vb_G = 0 \mod \ci$
by the first matrix and $\vb_F=\vc_F \mod \ci$ by the second one,
\ie\ ~$\vb=F^T\vc_F+I^I\vb_I$~ because the matrix elements can both be
non-vanishing only if $[\vb]\in\cf$ and $[\vc-\vb]\in\cg$.
Hence we are left with a sum over $\ci$ only.
Using again the notation of Appendix A we get now for the product
$$ \eqalign{
 \sum_{\vb_I\in\ci}  &\d^1\(F\1Q(a)+X_F(\vc_F+I_F^T\vb_I)\)  \cr
   &\d^1\(G\1Q(a)+GR(F^T\vc_F+I^T \vb_I)+
   Y_G(\vc_G-I_G^T\vb_I)\)\ ,
 \cr} \eqn\FGXY  $$
where $\1Q$ and $R$ are defined with respect to the basis of $\ch$,
whereas $X_F$ and $Y_G$ are given in the bases of $\cf$ and $\cg$,
respectively.
Equating the $\ci$-charges $IQ(a)$ as constrained by the two $\d$-functions,
we get a factor
$$ \d^1\(I_FX^T\vc_F+I_GY\vc_G-(X_I+Y_I-R_I)\vb_I\),\eqn\XYrestr $$
where $X_I=I_FXI_F^T$, $Y_I=I_GXI_G^T$ and $R_I=IRI^T$.
This is the analog of the last $\d$-function in \MXY\ and again determines
the subgroup $\ch'$ of allowed values $\vc$
(the ambiguity in the definition of $\vc_F$ and $\vc_G$ just amounts to
a shift in $\vb_I$ and thus is irrelevant, as it should be). Invertibility
of $\D_I=X_I+Y_I-R_I$ is again a sufficient condition for having the full
effective subgroup $\ch'=\ch$. If $\D_I$ is not invertible \XYrestr\
determines the subgroup $\ch'$ on which the product operates
non-trivially exactly as before. It consists of all the $\vc$'s for
which the $\delta$-restriction has a solution, but it seems hard to give
a more explicit description.

To show that the product is again of the form \DeltaForm\
we can now proceed just as in the previous case.
First we sum over those $\vb_I$ on which the summand doesn't
depend at all, namely $\vb_I \in \Ker X_F I_F^T  \cap \Ker Y_G^T I_G^T
= \Ker X \cap \Ker Y^T$. Then we solve (formally at least) \XYrestr\
which leaves us with a sum over $\Ker \D_I \cap (\Ker X
\cap \Ker Y^T)$. For each charge, there can be at most one
term in that sum that contributes.
Just as before, the sum
over $\vb_I \in \D_I$ restricts the first $\delta$-function to a
group $\ch'' \supset \ch'$.
To see this consider $\vnu \cdot \vec Q =
\vnu_F \cdot F\vec Q + \vnu_G \cdot G \vec Q$ for all $\vnu \in \ch'$.
Using \FGXY\ one can express this in terms of $\vc$ and $\vb$, and then
using \XYrestr\ one can show that the $\vb$-dependence cancels.
The rest is exactly as before,
because we now have
$|\ch/\ch'|=|\Im Y_I| / |\Im Y_I \cap \Im \D_I|$, which is equal to
$|\ch/\ch''|=|\Ker \D_I| / |\Ker \D_I \cap \Ker Y^T_I| $.

Thus, although we
could derive an explicit product formula only in case of equal subgroups,
we have been able to give a general proof of closure under
matrix multiplication.

\chapter{Discussion}

The central result of our investigation is that all simple current modular
invariants for arbitrary centers are of the form
$$ M_{a,[\vb]a} = \Mult(a) \prod_i \delta^1 ( Q_i (a) + X_{ij} \beta_j )
\eqn\Xformula \ ,$$
where the
simple currents $[\vb]$ are in a subgroup $\ch$ of the (effective) center and
where $R_\ch=X+X^T$ is the monodromy matrix for that subgroup with $X_{ij}N_j$
integer.
The number of these invariants is given by the sum of the number of
properly quantized antisymmetric matrices,
which is just the usual number of allowed discrete
torsions, over all subgroups of the center.
For symmetry groups that are generated by monodromy charges of
simple currents this shows that, in general,
the usual orbifold construction indeed gives all simple current invariants.
There is, however, a complication if there are odd (off-diagonal) entries in
the
monodromy matrix (for group factors of even order).
Then discrete torsion is a necessity rather than a possibility, an there is no
(genuine) orbifold with a modular invariant partition function.

Our procedure can be summarized as follows
\pointbegin We start with an orbifold inspired ansatz for a basis of
invariants.
\point We use the fact that (according to \BeBT)
$M_{a,[\vb]a}$ can only depend on the charge of $a$ (provided $S$  satisfies
the regularity conditions of \BeBT).
\point Fourier analysis of any such invariant shows that it can be expressed
as a complex linear combination of the basis of point 1.
\point Modular invariance and factorization restricts the complex
coefficients to certain phases. Modular invariance is used at one and two
loops. The two-loop transformation is an assumption, since for a general
CFT's we do not know explicit formulas for the two-loop characters and
their transformations.
\point The resulting invariants are positive. In all cases were a
comparison is possible, they coincide with the already classified
simple current invariants. In all cases where orbifolding works
straightforwardly, they coincide with the set of
orbifold invariants with
discrete torsion. In all other cases, it is in any case true that
for any subgroup $\ch$ of the center, after a redefinition of the phases
the remaining freedom is exactly
like the one for discrete torsions of orbifolds

The above formula is extremely useful in discussing some general features
of simple current invariants, which we did in section 3.
Obviously, the kernels of $X^T$ and $X$ define
the left and right extensions of the chiral algebras. Some conditions on
monodromies for heterotic combinations are immediate
from $X+X^T=R $. We
also showed that the quotients of the effective subgroup by the left/right
algebra extensions are isomorphic.
The chiral algebra is not enlarged
if and only if $X$ is invertible. Then $X$ defines
an automorphism and we recover the results of a classification of that type
of invariants \BeBe.
Finally we considered multiplication of our modular invariants, which
are functions of subgroups of the center and matrices $X$.
Although we did not present an explicit formula for arbitrary products
we did prove that our set of solutions closes under matrix
multiplication. This is an important check of the two-loop
assumption mentioned above.

Our results are also very useful from a practical point of view.
The structure of the final result is very simple so that one can easily
search for particular features, or implement a fast code for systematic
constructions of all different invariants.
Redundancies still could arise in case of permutation symmetries of
tensor products of identical conformal field theories, but these can now
straightforwardly be eliminated with the same methods as for orbifolds \ADE.
Along these lines, however, an important problem remains to extend our
classification to cases where exceptional invariants are know and, for
example, find a characterization of all invariants that can be written
as products of simple current invariants and the exceptional one(s).

\Appendix{A}

In this appendix we show how to write a modular invariant defined
with respect to a subgroup $\ch$ in terms of charges and currents
of a group $\cg \supset \ch$.

Consider a set of $N$ currents  $J_i$ generating a group $\cg$ and another
set of $M$ currents $J_a$ generating a subgroup $\ch$. Then there is
a $M \times N$ integral matrix
$H$ such that
$$ J_a = \prod_i  {J_i}^{H_{ai}} \eqn\Jrest $$
The charges $Q_a$ with respect to $J_a$ are related to the ${\cg}$-charges
$Q_i$ as
$$   Q_a = \sum_i H_{ai} Q_i. \eqn\Qrest $$
Currents in $\ch$ are of the form $[\va]=\prod_a {J_a}^{\a_a} $. In the
original basis one therefore has
$\a_i=\sum_a\a_a H_{ai}$.
Denoting charges $Q_a$ as $\vec Q_H$ and similarly for $Q_i$,
$\alpha_a$ and $\alpha_i$ we can summarize this as
$$\eqalign{ \vec Q_H &= H \vec Q_G \cr
            \va_G &= H^T \va_H \  . \cr }\eqn\JQrest $$
The $R$-matrix for the currents of $\ch$ is given by
$R_H=H R H^T$. An invariant
of the form \DeltaForm\ defined within $\ch$
and with a matrix $X_H$
can be written in the basis
of $G$ in the following way:
$$ M_{a,[\vb]a} = \delta^1 ( H \vec Q + X_H H^T \vb) \delta^1(U \vb) \ .
\eqn\Mrest $$
Here $\vec Q$ and $\vb$ are defined with respect to the
$\cg$-basis (note that in general there does not exist a representation
$X_H=HXH^T$ with a matrix $X$ satisfying the quantization conditions of $\cg$).
As usual in the $\delta$-functions a product over
all components of the $\cg$ basis is implicit.
The matrix $U$  is a rational matrix, quantized like $X$, with the
property that $\Ker U = \ch$. It ensures that $M$ has no
matrix elements related to currents that are not in $\ch$.

If we restrict $\vb$ to $\ch$ as required by the second $\d$-function we may
write the argument of the first $\delta$ as $
H \vec Q + X_H \vb_H = Q_H + X_H \vb_H$, which indeed is the correct
expression within $\ch$.

It is undoubtedly possible to prove
that any partition function of the form \Mrest\ is
one-loop modular invariant provided that
$\Im H = \Ker U \equiv \ch $ and that
$X_{H} + X_{H}^T = R_{H}$ (\ie\ no higher loop considerations
are required here). We only need this fact in cases where it is already
known that $\Ker U \subset \Im H$. To prove it, we make use of a result
of \BeBT\ that for any simple current invariant the sum over a row
(or column) of $M_{ab}$ is equal to zero
if the corresponding field is non-local with respect to the
left (or right) chiral algebra, and otherwise it is equal to the number of
currents in the chiral algebra. This implies that if we view
$M_{a, [\va]a}$ as a function of the charge rather than $a$, then
the sum over all currents and all allowed charges is always equal
to
$|{\ch}|$ for any group $\ch$ that contains all
currents $\va$ for which some $M_{a,[\va]a} \not = 0$. [To see this note
that it is manifestly true if there is no chiral algebra; if there is
a chiral algebra with $N$ currents, then the number of allowed charges
is reduced by a factor $N$, but each of the surviving charges now contributes
$N$ times to the sum].

Now the first factor in \Mrest\ satisfies this sumrule for any $X_H$. If
$\Ker U$ is smaller than $\Im H$ this means that the second
$\delta$ imposes extra restrictions on $\ch$-currents, so that
some matrix elements are put to zero. But then clearly the sum rule
is not satisfied for $\ch$ and the result cannot be modular invariant.

If one makes the ansatz \Mrest\ with $\Im H = \Ker U$,
the restriction on $X$ follows already from $T$-invariance alone.
If $M_{a, [\vb]a}$ is non-zero, it must be true that
$h(a) = h ([\vb]a) \mod 1$. This yields the condition
$h(\vb) = \vb \cdot \vec Q(a) \mod 1$, which within $\ch$
(and with $\ch$-indices omitted) can be written as
$$ -\half \vb \cdot R\cdot \vb = - \vb \cdot X \cdot \vb
\ \mod 1 \ . \eqn\RXrel$$
This must be true for any $\vb$ since there is always  a charge
$-X\vb$ for which $M_{a,[\vb]a}$ is non-zero. It is then clear that \RXrel\
is equivalent to $X + X^T = R $, with the usual periodicities.

\par \penalty-4000\vskip\chapterskip
   \spacecheck\referenceminspace \immediate\closeout\referencewrite
   \referenceopenfalse
   \line{\fourteenrm\hfil REFERENCES\hfil}\vskip\headskip
   \endlinechar=-1
   \input referenc.texauxil
   \endlinechar=13
   
\end